\documentclass[10pt, a4paper]{article}
\usepackage{lt4all}
\usepackage{graphicx}
\usepackage{tabularx}
\usepackage{soul}

\usepackage{epstopdf}
\usepackage[latin1]{inputenc}

\usepackage{hyperref}
\usepackage{xstring}

\title{Talking with Robots: Opportunities and Challenges}

\name{Roger K. Moore}

\address{Speech and Hearing Research Group (SPandH) \\
	University of Sheffield \\
	Regent Court, 211 Portobello \\
	Sheffield, S1 4DP, UK \\
         r.k.moore@sheffield.ac.uk \\}

\abstract{
Notwithstanding the tremendous progress that is taking place in spoken language technology, effective speech-based human-robot interaction still raises a number of important challenges.  Not only do the fields of robotics and spoken language technology present their own special problems, but their combination raises an additional set of issues.  In particular, there is a large gap between the formulaic speech that typifies contemporary spoken dialogue systems and the flexible nature of human-human conversation.  It is pointed out that grounded and situated speech-based human-robot interaction may lead to deeper insights into the pragmatics of language usage, thereby overcoming the current `habitability gap'.  \\ \newline  \\ \newline
\centerline{\bf Résumé}  \newline Malgré les énormes progrès réalisés dans la technologie de la langue parlée, une interaction homme-robot efficace basée sur la parole soulève encore un certain nombre de défis importants. Non seulement les domaines de la robotique et de la technologie de la langue parlée posent des problèmes particuliers, mais leur combinaison soulève un ensemble de problèmes supplémentaires. En particulier, il existe un large fossé entre le discours stéréotypé qui caractérise les systèmes de dialogue parlés contemporains et la nature flexible de la conversation homme-humain. Il est souligné que l'interaction homme-robot fondée et basée sur la parole peut mener à une compréhension plus approfondie de la pragmatique de l'utilisation du langage, surmontant ainsi le `fossé d'habitabilité' actuel.}

\begin{document}

\maketitleabstract

\section{Introduction}

\begin{figure}[!h]
\begin{center}
\includegraphics[width=\columnwidth]{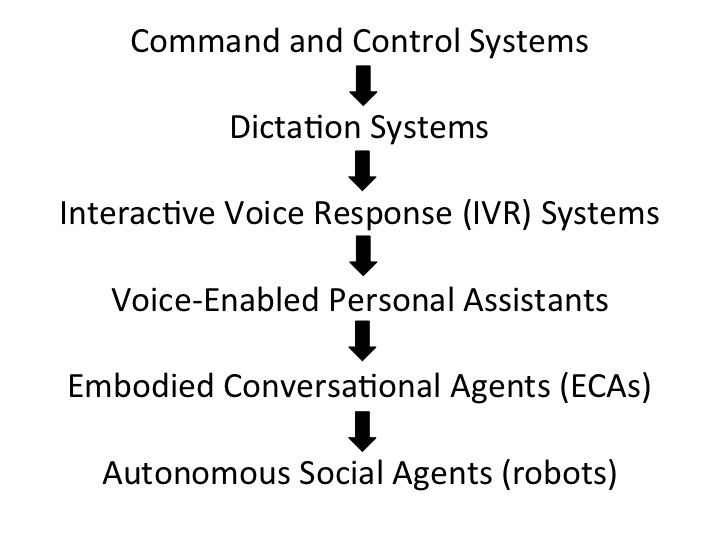} 
\caption{The evolution of spoken language processing applications from specialised military `command-and-control' systems of the 1980/90s to contemporary 'voice-enabled personal assistants' (such as \emph{Siri} and  \emph{Alexa}) and future `autonomous social agents', i.e.\ robots.}
\label{fig:DEV}
\end{center}
\end{figure}

Recent years have seen tremendous progress in the deployment of practical spoken language systems - see Figure~\ref{fig:DEV}.  Commencing in the 1980s with the appearance of specialised isolated-word recognition (IWR) systems for military command-and-control equipment, spoken language technology has evolved from large-vocabulary continuous speech recognition (LVCSR) for dictating documents (such as Dragon's \emph{Naturally Speaking} and IBM's \emph{Via Voice}) released in the late 1990s, through telephone-based interactive voice response (IVR) systems to the launch of \emph{Siri} (Apple's voice-enabled personal assistant for the iPhone) in 2011.  \emph{Siri} was quickly followed by Google \emph{Now} and Microsoft's \emph{Cortana}.  The following years heralded a new era of smart speaker based voice assistants, starting with  Amazon's 2015 release of \emph{Alexa} followed later by \emph{Google Home}, Apple's \emph{HomePod} and Sonos \emph{One}.  

These contemporary systems not only represent the successful culmination of over 50 years of laboratory-based speech technology research \cite{Pieraccini2012}, but also signify that speech technology had finally become ``mainstream'' \cite{Huang2002} (at least, in the English-speaking world).  Indeed, the market penetration of these smartphone and smart speaker based voice assistants is astounding.  For example, Siri has had over 40 million monthly active users in the U.S.\ since July 2017, Google Assistant is available on over 225 home-control brands and more than 1,500 devices, and tens of millions of Alexa-enabled devices were sold worldwide over the 2017 Christmas holiday season \cite{Boyd2018}.  Also, a study by Juniper Research \cite{Juniper2017} estimated that the number of voice assistant devices across all platforms (smartphones, tablets, PCs, speakers, connected TVs, cars and wearables) would reach 870 million in the U.S.\ by 2022.  

Research is now focused on verbal interaction with embodied conversational agents (such as on-screen avatars) and autonomous social agents (such as robots), based on the assumption that spoken language will provide a `natural' interface between human beings and future (so-called) intelligent systems, and first-generation devices (such as \emph{FurHat}\footnote{\url{https://www.furhatrobotics.com}} and \emph{Olly}\footnote{\url{https://www.heyolly.com}}) have already begun to enter the commercial marketplace.  

However, notable casualties (such as \emph{Jibo}\footnote{\url{https://www.jibo.com}} which famously announced its own demise in June 2019) confirm that there are significant challenges as well as opportunities in creating spoken language based interaction between people and robots \cite{Moore2015}.  Some of these are discussed below.

\section{Why Robots?}

Before discussing the challenges of talking with robots, it is useful to recall why robots are of interest in the first place.  First and foremost, developments in robotics are driven by the many benefits provided by \emph{automation}.  Since the beginning of time, humans have been inventing technologies to ease their daily toil, and the industrial revolution heralded an era of increasing automation using ever more sophisticated machines.  The benefits of doing so include making/saving money, saving time and effort and improving the quality of life.  Robotics - driven by the recent surge in artificial intelligence (AI) - represents the latest attempts at automation, particularly for doing things that are difficult, dirty, dangerous or dull.

\section{What is a Robot?}

A robot is harder to define that one might think.  As Joseph Engelberger (1925-2015), developer of the first industrial robot in the United States in the 1950s, famously said: ``\emph{I can't define a robot, but I know one when I see one}''!

In fact there are a number of definitions of a robot, and the following is typical \ldots
\begin{quotation}
``\emph{A robot is an actuated mechanism programmable in two or more axes with a degree of autonomy, moving within its environment, to perform intended tasks.}''\footnote{\url{http://www.leorobotics.nl/definition-robots-and-robotics}}
\end{quotation}

They key idea is that a robot is a \emph{physical machine} (i.e.\ capable of movement within in environment, whether it is real or simulated), \emph{autonomous} (i.e.\ capable of acting without constant human intervention) and \emph{programmable} (i.e.\ it is more than just an automaton).  This means that Siri and Alexa are \emph{not} robots (since they are incapable of moving or acting on the world), nor are tele-operated devices such as remote-controlled drones (since they are not autonomous), and nor is Terminator (since it is purely fictional!).  Typical robots are thus those that one would find on an industrial production line, floor-cleaning robots (such as Roomba\footnote{\url{https://www.irobot.co.uk/roomba}}), and humanoid robots (such as Pepper\footnote{\url{https://www.softbankrobotics.com/emea/en/pepper}}).

\section{Why Talk with a Robot?}

As with all technology, there are huge benefits to be gained when humans are `in the loop'.  For example, a modern automobile already exhibits several levels of automation (e.g.\ power-assisted steering and cruise control) combined with human involvement in low-level activities such as acceleration and braking.  As technology moves towards more autonomous vehicles and the degree of automation increases, human involvement will shift to higher levels (such as defining the destination and required time of arrival) with low-level interventions only occurring in exceptional circumstances (e.g.\ in an emergency).  Such high-level interactions would seem to be very appropriate for a communication channel such as speech.

The field of `human-robot interaction' (HRI) is concerned with these issues and, in particular, how to maximise the effectiveness of such interaction in a multi-modal context, e.g.\ vision, sound, haptics, and of special interest here, speech and language.  So, how might spoken language play a role in human-robot interaction?  This can be answered by considering three domains in which such interaction might take place: the \emph{physical} world of stuff and things, the \emph{social} world of people, agents and relations, and the \emph{abstract} world of ideas, information, data and thought.

\subsection{Speech-based HRI in the Physical World}

Human-robot interaction in the physical world is often concerned with the provision of mechanical support for the human being, e.g.\ allowing a person to lift a heavy object or pilot a vehicle.  Much of the low-level interaction could be achieved by the manual operation of physical controls and observing visual displays, but the introduction of a speech channel would facilitate additional control even if the users hands are occupied, and/or the ability to receive information even if the eyes are engaged in a more critical task (such as watching for hazards).  Such activities are known as \emph{eyes-busy, hands-busy} scenarios, and they are prime candidates for speech-based HRI.

In general, physical HRI is targeted at \emph{collaborative working} where tasks are distributed between human and robot teams.  In such situations, speech can offer a powerful means for coordinating actions (``\emph{Pass me the wrench}'') and for managing joint attention (``\emph{Mind that hole!}'').

\subsection{Speech-based HRI in the Social World}

Human-robot interaction in the social world is concerned with the provision of emotional and/or motivational support for the human being, e.g.\ through \emph{companionship} and the exhibition of empathy or even dominance (as would be required from a personal trainer).  Such behaviours would serve to underpin the relations between the different actors/agents and their individual and/or collective roles and responsibilities.  

In general, social HRI would exploit both verbal and non-verbal channels of communication, and would naturally draw on the expressive \emph{paralinguistic} properties of spoken language.

\subsection{Speech-based HRI in the Abstract World}

Human-robot interaction in the abstract world is concerned with the provision of mental support for the human being, e.g.\ by giving access to the vast amounts of information/data available on the internet.  Spoken language not only offers a more intuitive (some say `natural') method of human-robot communication, but it also supports a very high information-rate exchange compared to that available through the physical or social channels.

\section{Challenges for Speech-based HRI}

\subsection{Issues Arising from Robotics}

There are many challenges facing the opportunities identified above.  Not only are there a number of difficulties to be overcome in the core area of speech-based human-robot interaction, but problems are also inherited from the field of robotics in general.  For example, all robots are complex mechanical, electrical, electronic and computer-based physical machines operating in the real world, which means that they can be very fragile.  A network outage, a broken spring, or a computer bug can easily bring operations to a halt (or worse), and the likelihood of some component failing can be quite high.  Also, robots tend to be quite expensive pieces of equipment, meaning that personal ownership may be challenging for particular user groups.

\subsection{Issues Arising from Spoken Language Technology}

Likewise, all the problems facing mainstream spoken language technology also apply to speech-based human-robot interaction.  For example, strong accents, minority languages, and noisy environments can all lead to poor performance of the speech technology components which, in turn, will have a negative impact on the effectiveness of speech-based HRI.

\subsection{Issues Arising from Speech-based HRI}

In addition, there are many issues that arise from speech-based human-robot interaction itself.  For example, robots are quite noisy, hence listening and moving are often incompatible activities\footnote{One well known robot even has its microphones mounted immediately adjacent to its cooling fans!}!  Also, everyday environments may contain many individuals (and maybe many robots).  So figuring out who is where, isolating an individual from a crowd, knowing whether one is being addressed, or timing an intervention in an ongoing conversation all present major difficulties that require beyond state-of-the-art solutions.

\begin{figure}[!h]
\begin{center}
\includegraphics[width=\columnwidth]{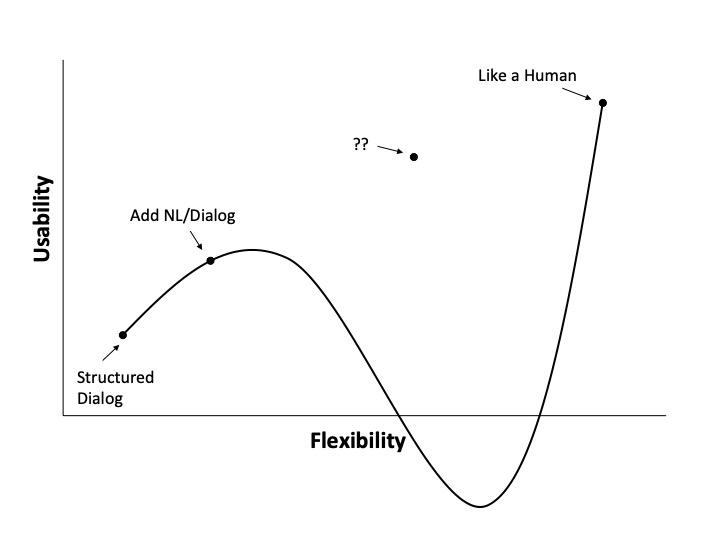} 
\caption{Illustration of the drop in usability that can occur in a spoken language dialogue system when its flexibility is increased.}
\label{fig:MP}
\end{center}
\end{figure}

Even if some of these practical problems could be overcome, there are still issues concerning the role of \emph{language} in human-robot interaction.  For example, studies into the usage of smart assistants suggest that, far from engaging in a promised natural `conversational' interaction, users tend to resort to formulaic language and focus on a handful of niche applications which work for them \cite{Moore2016a}.  Given the pace of technological development, it might be expected that the capabilities of such devices will improve steadily, but according to \newcite{Philips2006} there is a `habitability gap' in which usability drops as flexibility increases - see Figure~\ref{fig:MP}.

\begin{figure}[!h]
\begin{center}
\includegraphics[width=\columnwidth]{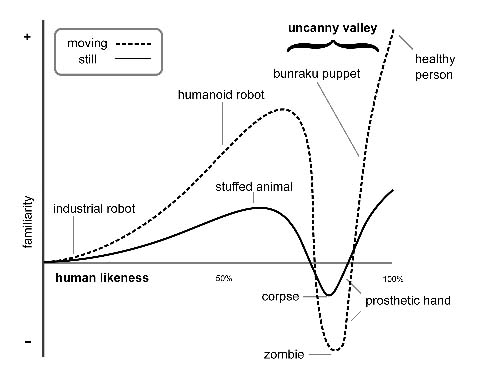} 
\caption{Illustration of the `uncanny valley' effect in which a near human-looking artefact (such as a humanoid robot) can trigger feelings of eeriness and repulsion.}
\label{fig:UV}
\end{center}
\end{figure}

It has been hypothesised that the habitability gap is a manifestation of the `uncanny valley' effect (see Figure~\ref{fig:UV}) whereby a near human-looking artefact (such as a humanoid robot) can trigger feelings of eeriness and repulsion \cite{Mori1970}.  In particular, a Bayesian model of the uncanny valley effect \cite{Moore2012} reveals that it can be caused by \emph{misaligned} perceptual cues.  Hence, a device with an \emph{inappropriate} voice can create unnecessary confusion in a user.  For example, the use of human-like voices for artificial devices encourages users to overestimate their linguistic and cognitive capabilities.

The Bayesian model of the uncanny valley effect suggests that the habitability gap can only be avoided if the visual, vocal, behavioural and cognitive \emph{affordances} of an artefact are aligned.  Given that the state-of-the-art in these areas varies significantly, this means that the capabilities of an artificial agent should be determined by the affordance with the lowest capability \cite{Moore2017,Wilson2017}.  In other words, emulating a human is a recipe for failure, rather ``\emph{it is better to be a good machine than a bad person}'' \cite{Balentine2007}.

Another significant shortfall in our current level of knowledge about creating effective speech-based human-robot interaction is that robots need to \emph{understand}, not just speak and listen.  This is already a major impediment to conversational interaction with contemporary smart assistants.  However, there is hope that deeper insights into the problem may arise from tackling language-based HRI on the basis that such interaction is necessarily \emph{situated} and \emph{grounded}; both of which are considered to be key aspects of genuine language understanding and give support to the `pragmatics-first' view of language \cite{Bar-On2017a}.

\subsection{Ethical Issues}

Finally, the drive towards speech-based human-robot interaction also raises a number of important ethical concerns.  For example, the appearance of smart assistants in people's homes has already sparked controversy about whether such devices are listening to private conversations and sending sensitive personal information to unidentified third-parties.  As a result, the level of \emph{trust} that a user can place in an artificial conversational partner has become a subject of much debate.

Another area of concern is the ability to \emph{fake} abilities that are far beyond the state-of-the-art.  There are already examples of so-called `intelligent' conversational robots being demonstrated to the public and the press which, on investigation, turned out to be operated by human beings, either remotely or even inside an elaborate robot costume!  Such unethical activities tend to fuel the technological \emph{hype} that often surrounds robots and speech-based interaction with them.  Preprogrammed spoken responses to scripted verbal questions are easy to arrange, but at best seriously misrepresent the actual capabilities of the the device, and at worst undermines the confidence of funding agencies in determining what research (if any) needs to be supported.

\section{Conclusion}

Notwithstanding the tremendous progress that is currently taking place in spoken language technology, the achievement of effective speech-based human-robot interaction still raises a number of important challenges.  Not only do the two fields of robotics and spoken language technology present their own special problems, but their combination raises an additional set of issues that are worthy of investigation.  In particular, it is noted that there is a large gap between the type of formulaic speech-based interaction that typifies contemporary spoken language dialogue systems and the fully flexible natural language interaction exhibited in human-human conversation \cite{Moore2016b}.  Nevertheless, it is pointed out that the grounded and situated nature of speech-based human-robot interaction may lead to deeper insights into the pragmatics of language usage in real-world environments, thereby overcoming the current `habitability gap'.

\section{Bibliographical References}

\bibliographystyle{lt4all}
\bibliography{Moore-TalkingwithRobots}

\end{document}